\documentstyle[aps]{revtex}
    \def \be   {\begin{equation}}
\def \ee   {\end{equation}}
\def \l {\label}
\begin{document}
\input epsf
\title{DISCRETE AND FINITE GENERAL RELATIVITY }
\author{Manoelito M de Souza and Robson N. Silveira}
\address{Universidade Federal do Esp\'{\i}rito Santo - Departamento de
F\'{\i}sica\\29065.900 -Vit\'oria-ES-Brasil}
\date{\today}
\maketitle
\begin{abstract}
 We develop the General Theory of Relativity in a formalism with extended causality that describes physical interaction through discrete, transversal and localized point-like fields. The essence of this approach is of working with fields defined with support on straightlines and not on hypersurfaces as usual. The homogeneous field equations of General Relativity are then solved for a finite, singularity-free, point-like field that we associate to a ``classical graviton". The standard Einstein's continuous formalism is retrieved by means of an averaging process, and its continuous solutions are determined by the chosen imposed symmetry. The Schwarzschild metric is obtained by imposing spherical symmetry on the averaged field.
\end{abstract}
\begin{center}
\end{center}

\section{Introduction}
\noindent If we can assume that the gravitational interaction between two masses, as any other elementary interaction, is fundamentally of a quantum nature, that is, mediated by  a discrete and localized agent (the graviton), then the General Theory of Relativity (GR ) is a wonderful  average  geometric description of such phenomenon: it replaces this intermediating agent by the metric of a curved spacetime, smoothing and hiding the discreteness, the localized and anisotropic  aspects of the quantum interaction; it is a wonderful description because it fits \cite{Will} our observational data, which can be seen then as a consequence of the minuteness of the action of a single graviton. Thus, GR  just describes an effective  average continuous interaction. It would not be correct to try to reobtain the fundamental quantum picture from this effective description, and besides, this would be impossible as they have support on distinct and not-compatible topologies. The concept of a corpuscle  (a quantum) is associated to the geometry of a line (its support manifold, free of singularity) while the field interaction of GR  is associated to the geometry of a lightcone, which is singular at its vertex. This may explain the singularities of GR  and the difficulties with its quantization. 

Another procedure \cite{hep-th/9610028,hep-th/9708066,hep-th/9712069}, that in our opinion is the most appropriate, would be, back to the origins, start with a classical description that contemplates the discrete, localized and anisotropic nature of the fundamental interactions, and that, upon an averaging procedure, reproduces GR  as a theory of continuous and distributed field.  Actually, this should apply to
  all fundamental-field theories.  Positive results  have been obtained  with a such approach on Electrodynamics \cite{hep-th/9712069}, and now we describe our first similar experiences with GR.  Frequently we shall refer to this previous work for comparison and for stressing the similarities between the electromagnetic and the gravitational fields. The gravitational radiation, as a pulse generated by a point-source and described in GR  by the metric field $g_{\mu\nu}$, with support on the lightcone, is replaced, in our approach, by a point-like metric field $g^f_{\mu\nu}$, with support on a straight line, a lightcone generator, tangent to a null four-vector $f$, $f^2=0$. $g^f_{\mu\nu}$ can be pictured as the intersection of $g_{\mu\nu}$ with the lightcone generator $f$. The physical idea is that this point field propagating on a lightcone generator represents a ``classical graviton", a local deformation on the otherwise Minkowski background. The straight line tangent to $f$ is a generic one, determined exactly by the presence of a classical graviton on it, which breaks the spherical symmetry, the space isotropy around the point-like source, but without ever breaking the manifest (Lorentz) covariance.  This is realized with the concept of extended causality \cite{hep-th/9708066}, which for completeness we review in Section II. The contact with standard GR  is exactly that the straight-line support of $g^f_{\mu\nu}$ is a generator of the lightcone, the support of $g_{\mu\nu}$. Thus an integration over the cone generators $f$ of $g^f_{\mu\nu}$ reproduces $g_{\mu\nu}$ as an average-valued field. The chosen symmetry determines the average solution.  It turns out that all standard GR  solutions, not only the radiation ones, can in principle, be retrieved as  such averages. A spherical symmetry for a vacuum solution reproduces the Schwarzschild metric. This is particularly interesting because it implies on the possibility of regarding even a static solution as the average or the effective result of a radiation field.
 The Vaidya's metric \cite{Vaidya}, as an example of non-vacuum solution, is retrieved in a similar way from the discrete solutions of an spherical distribution of point-like massless sources \cite{inpreparation}.\\
Although this paper is about a new approach to classical field theory, GR in particular, its main motivations come, nonetheless, from the dream of a finite and consistent quantum field theory for all fundamental interactions. It is even written with an eye on a  subsequent quantization step.
It is well known that both classical and quantum field theories are plagued with problems of infinities, locality and causality violations. Classical Electrodynamics for example, our best paradigmatic classical field theory, is not completely consistent because its fields diverge when taken over their point-like sources. This is sometimes erroneously attributed to the assumption of a point-like source. It is an old problem that has resisted for over a century  the most varied and persistent efforts of searching solutions. For persisting and for just being aggravated in a quantum theory, it is now considered for many as just an indication of the inadequacy of our pseudo-Riemannian model of  spacetime. According to this vision, in the zero-distance limit the spacetime should lose some of its assumed properties like continuity, or commutativity, or simply it should not exist as such in this limit. These are all radical proposals that show how deep is the actual crisis in field theory.  It has been shown in reference \cite{hep-th/9610028} that this does not need to be the case. It is not necessary any change in the spacetime structure nor on the Maxwell's equations; it is just a matter of better understanding the physical picture. If an appropriate zero-distance limit is correctly taken, the solutions to Maxwell's equations for the field of a  point-like electron
are free of these infinities and of causality-violating problems in the electron equation of motion. The price to be paid is the anticipated recognition of the discrete (quantum) character of the process of emission/absorption of light by the electron, i.e. the anticipation of the Planck-Einstein concept of photon to Classical Electrodynamics on its zero-distance limit. It calls for a revision of our ideas about the physical meaning of a field, of its singularities, and of the equations that describe its evolution. The classical Maxwell field must be seen as a spacetime average (over the lightcone) of these discrete emitted/absorbed fields. A finite and consistent \cite{hep-th/9712069} classical theory of light is defined by the Maxwell's equations, formulated in terms of discrete (defined on the lightcone generators) fields. Is this idea that we want to apply here in GR.\\
Producing a finite classical theory is highly desirable but is not sufficient to assure that it will remain finite after being quantized. A quantum theory has further infinities that needs further renormalizations. Although a renormalization process can make sense of the perturbative series expansion, notwithstanding lingering questions \cite{Landau,Wilczek} concerning its convergence and the theory very existence for all but the trivial non-interacting case, for gravitation there is no doubt  about its total failure.  
There are infinitely many ways of decomposing a continuous and distributed field in terms of discrete elements if they do not have a pre-fixed energy-moment content. These are the well known causes of infrared divergence in a quantum theory. They should not appear if one had started from discrete point-like fields with a previously fixed energy-moment content. Being on a lightcone generator is a fundamental  feature in this new approach because it fixes both the energy and the moment of the point-like field which eliminates the infrared divergences, and prohibits virtual off-shell and acausal interactions that generate ultraviolet divergences. This new  approach generates a finite classical field theory, as we will show in the following, and we can expect then that this desirable property must remain after the theory quantization.\\
Quantization will not be our subject here but this approach to GR is obviously relevant to quantum gravity, to the nature and meaning of singularity as $g^f_{\mu\nu}$ is not singular while its lightcone average $g_{\mu\nu}$, regardless its symmetry, is singular at the lightcone vertex. Also relevant to field theory is that in order to retrieve a Coulombian-type of field in this lightcone averaging process, it must necessarily include non-physical longitudinal excitations. This happens to the electromagnetic field \cite{hep-th/9712069} and,  basing on some known theorems\cite{Strochi}, it can be expected happening also with generic (classical and quantum) non-abelian fields.

In Section II, for the sake of completeness, we reproduce a brief review \cite{hep-th/9708066,hep-th/9712069} of extended causality and its applications to field theory. In Section III we show how the discrete fundamental field can be seen as an elementary part  of the standard continuous field. The second field is not necessary for defining the first one; this is just an heuristic view. The theory must be defined in terms of the discrete field; the continuous one and its standard formalism are retrieved in terms of effective averages of the discrete field.  
The General Theory of Relativity, in terms of discrete fields, is described in Section IV, and in Section V the homogeneous field equations are solved for a discrete solution. The Schwarzschild metric is recovered in Section VI, with the assumption of spherical symmetry. Finally we conclude, in Section VII, discussing its physical meaning and implications.

\section{Causality in Field Theory}

\noindent As we want to describe a {\it free} massless point object moving on a straight-line between two successive discrete interaction events on a Minkowski background manifold we have to impose on its propagation two constraints  that describe, respectively, its lightcone and its tangent hyperplane, in order to covariantly define its straightline support, a lightcone generator. We associate these constraints to the idea of causality. Actually, in this section we present a more generic formalism that is valid for massive fields too. 

\noindent Any given pair of events on Minkowski spacetime defines a four-vector $\Delta x.$ If this  $\Delta x$ is connected to the propagation of a physical object (a signal, a particle, a field, etc) it is constrained to
\be
\label{1}
\Delta\tau^2=-\Delta x^{2}.
\ee 
Our metric is $\eta=diag(1,1,1,-1)$ and in our notation we omit the spacetime indices when this does not compromise the text comprehension. So, $x$ stands for $x^{\mu},$ $\; \partial$ for $\partial_{\mu}$, and $A(x,\tau)$ for a vector field $A^{\mu}(x,\tau)$, for example. $\tau$ is a real-valued parameter. So, (\ref{1}) just expresses that $\Delta x$ cannot be spacelike. A physical object does not propagate over a spacelike $\Delta x.$ This is {\it local causality}, and (\ref{1}) defines the change of propertime $\Delta\tau$ associated to $\Delta x.$  Geometrically it is the definition of a three-dimensional double cone; $\Delta x$ is the four-vector separation between a generic event $x^{\mu}\equiv({\vec x},t)$ and the cone vertex. See the Figure 1. This conic hypersurface, in field theory, is the free-field support: a free field cannot be inside nor outside but only on the cone. The cone-aperture angle $\theta$ is given by
\be
\tan\theta=\frac{|\Delta {\vec x}|}{|\Delta t|},\qquad c=1,
\ee
or $\Delta\tau^{2}=(\Delta t)^{2}(1-\tan^{2}\theta).$
A change of the supporting cone corresponds to a change of speed of propagation and is an indication of interaction.
Special Relativity restricts $\theta$ to the range $0\le\theta\le\frac{\pi}{4},$ which corresponds to a restriction on $\Delta\tau:$ $0\le|\Delta\tau|\le|\Delta t|.$ The lightcone ($\theta=\frac{\pi}{4},$ or $|\Delta\tau|=0$) and the t-axis in the object rest-frame ($\theta=0,$ or $|\Delta\tau|=|\Delta t|$) are the extremal cases. 

\mbox{}

\vglue-2cm
\begin{minipage}[]{5.0cm}\hglue-3.50cm
\parbox[]{5.0cm}{
\begin{figure}
\epsfxsize=400pt
\epsfbox{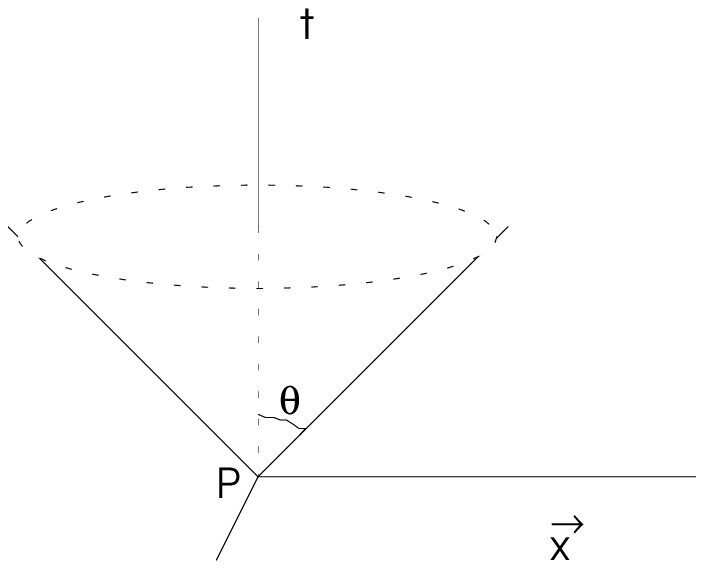}
\vglue-4cm
\end{figure}}\\
\end{minipage}\hfill
\hfill 
\mbox{}

\hspace{8.0cm}
\vglue-12cm
\begin{minipage}[]{7.0cm}\hglue7.0cm
\parbox[b]{7.0cm}
{\vglue-17cm \hspace{5.0cm}\parbox[t]{7.0cm}{
Fig.1. The relation $\Delta\tau^2=-\Delta x^{2},$ a causality constraint, is seen as a restriction of access to regions of spacetime. It defines a three-dimension cone which is the spacetime available to a point, free, physical object at the cone vertex. The object is constrained to be on the cone.}}\\ \mbox{}
\vglue4cm
\end{minipage}

\vglue1cm
\noindent The concept of extended causality corresponds to a more restrictive constraint; it requires that (\ref{1}) be also applied to $x+dx$, an event in the same cone, in the neighbourhood of $x$, and for which we can write (\ref{1}) as $$(\Delta\tau+d\tau)^2=-(\Delta x+dx)^{2},$$ or just $\Delta\tau d\tau+\Delta x.dx=0,$ after making use of (\ref{1}). This is equivalent to the imposition of a second constraint, besides the first one (\ref{1}):
\be
\l{f}
d\tau+  f.dx=0.
\ee
$f$ is a constant, timelike $(f^{2}=-1$) four-vector tangent to the cone, a cone generator, and is defined by 
\be f^{\mu}=\frac{\Delta x^{\mu}}{\Delta\tau},
\ee
if $\Delta\tau\ne0;$ it is lightlike $(f^{2}=0$) in the limiting case when $\Delta\tau=0$. \\
The equation (\ref{f}) can be obtained from direct differentiation of (\ref{1}),  and geometrically it defines a hyperplane tangent to the cone (\ref{1}). Therefore, the simultaneous imposition of (\ref{1}) and of (\ref{f}) restricts the field support to the cone generator tangent to $f$, intersection of the hypercone (\ref{1}) and its tangent hyperplane (\ref{f}). According to (\ref{f}) 
\be
\l{del} 
f_{\mu}=-\frac{\partial\tau}{\partial x^{\mu}}.
\ee 
For $\Delta\tau=0,\;$ $f_{\mu}$ is orthogonal to the hyperplane (\ref{f}), but, at the lightcone vertex, it is also a lightcone generator.

 Imposing in field theory the two constraints, (\ref{1}) and (\ref{f}), instead of just (\ref{1}), as it is usually done, corresponds to knowing the initial position and velocity in point-particle dynamics.
One can summarise it by saying that while the local causality restricts the available space-time of a free physical object to a conic three-dimensional hypersurface, the extended causality restricts it to just a straight line, a cone generator.

\noindent For the propagation of a free object, $\Delta x$ and $dx$ are collinear. Together, the constraints (\ref{1}), written as $d\tau^2=-d x^{2},$ and (\ref{f}) are equivalent to the single condition $dx^{2}+(f.dx)^{2}=0$, that may be put as
\be
\l{lambda}
dx.\Lambda^{f}.dx=0,
\ee
with 
\be
\Lambda^{f}_{\mu\nu}=\eta_{\mu\nu}+f_{\mu}f_{\nu},
\ee
($f_{\mu}=\eta_{\mu\nu}f^{\nu}$),  which is a projector orthogonal to  $f^{\mu}$,  $  f.\Lambda.f=0.$  Therefore the constraint (\ref{lambda})   allows only displacements $dx^{\mu}$  parallel to $f^{\mu}.$ The eq. (\ref{lambda}) is useful for a more compact notation.\\
\noindent We should observe that the formalism presented in this section is specifically appropriate for solving homogeneous field equations as we are considering just the propagation of a field without mentioning its sources. The event at the cone vertex is kept fixed, and $\tau$ and $x$ are parameters of a same field. For solving a field equation with sources, as done in \cite{hep-th/9712069}, $\tau$ and $x$ in (\ref{f}) are parameters of two distinct objects, the electron and its self-field, respectively. The great difference is that, in the case of the field and its source, the gauge-fixing condition over the (massless) field fixes that the direction ${\vec f}$ of the emitted photon by an accelerated electron is orthogonal to the electron acceleration ${\vec a}$, on its instantaneous rest-frame at its retarded time:
\be
\l{fa}
{\vec a}.{\vec f}{\Big |}_{{{\vec V}=0}\atop{d x.\Lambda^{f}.d x=0}}=0.
\ee
This condition is enough to assure the field transversality. In the homogeneous case, treated here, we lose this information.
  
\section{Fields and field equations}

As a consequence of the causality constraint (\ref{1}), the fields must be explicit functions of x and of $\tau,$ where $\tau$, is a supposedly known function of x, a solution of (\ref{1}):  $$\tau=\tau_{0}\pm\sqrt{-(\Delta x)^{2}}.$$ 
 For a massless field, as it propagates without a change on its proper time, $\Delta\tau=0$, $\;\tau$ is actually the instantaneous proper-time of its source at the event of its emission.  Well-known examples of this are the Lienard-Wiechert solutions. See the Figure 2 where $z(\tau)$ is the source worldline parameterized by its proper time $\tau.$ 

\mbox{} 
\hglue-1.0cm
\begin{minipage}[]{5.0cm} 
\parbox[b]{5.0cm}{
\begin{figure}
\vglue2cm
\epsfxsize=100pt
\epsfbox{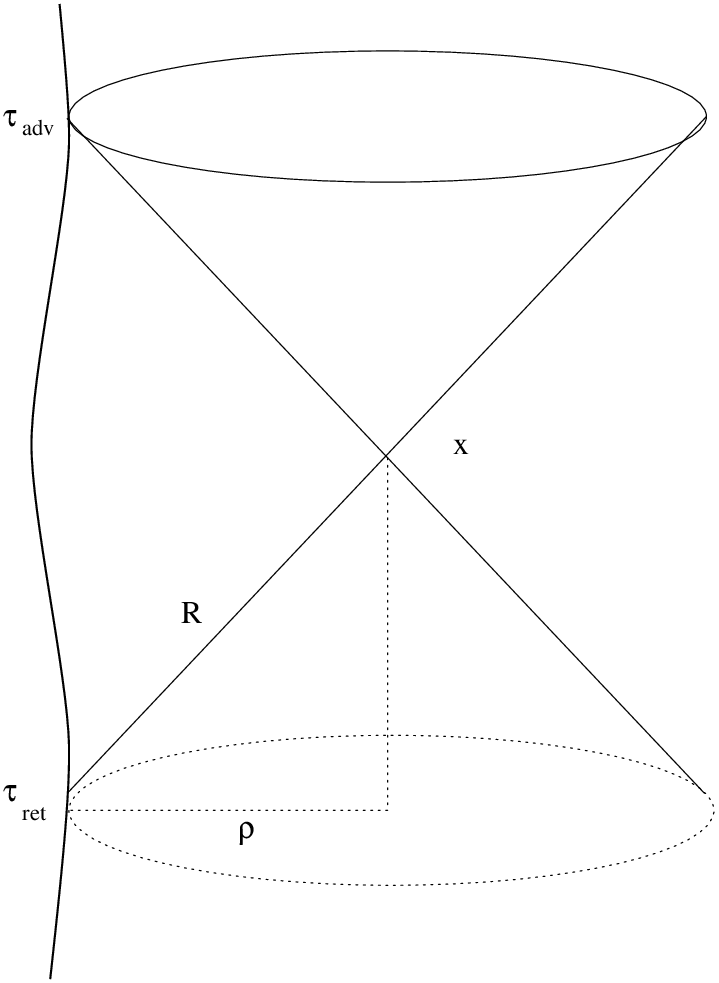}
\vglue-4cm
\end{figure}}\\
\end{minipage}\hfill
\hfill 
\mbox{}

\hspace{8.0cm}
\vglue-1cm
\begin{minipage}[]{7.0cm}\hglue7.0cm
\parbox[b]{7.0cm}
{\vglue-17cm \hspace{5.0cm}\parbox[t]{7.0cm}{
Fig.2. { 
{The usual interpretation of the Lienard-Wiechert solutions. By the point x passe two spherical waves: the retarded one, created in the past $\tau_{ret}$, and the advanced one, created in the future $\tau_{adv}.$ J is the source of both.}}}}
\vglue3cm
\end{minipage}

\vglue4cm

\noindent We turn now to the question of how to define a field with support on a generic fibre $f$, a $(1+1)$-manifold embedded on a $(3+1)$-Minkowski spacetime.
Let $A_{f}(x,\tau)$ be a $f$-field, that is, a field defined on a fibre   $f$. It is distinct of the field $A(x,\tau)$ of the standard formalism, which is defined on the cone. $A_{f}(x,\tau)$ may be seen as the restriction of $A(x,\tau)$ to a fibre $f$,  
\be
\l{Af}
A(x,\tau)_{f}=A(x,\tau){\Big |} _{\Delta x.\Lambda^{f}.\Delta x=0}
\ee
It is a point-like field, the intersection of the wave-front $A(x,\tau)$ with the fibre $f$. See the Figure 3.

\vglue-2cm

\vglue1cm
\parbox[]{5.0cm}{
\epsfxsize=400pt
\epsfbox{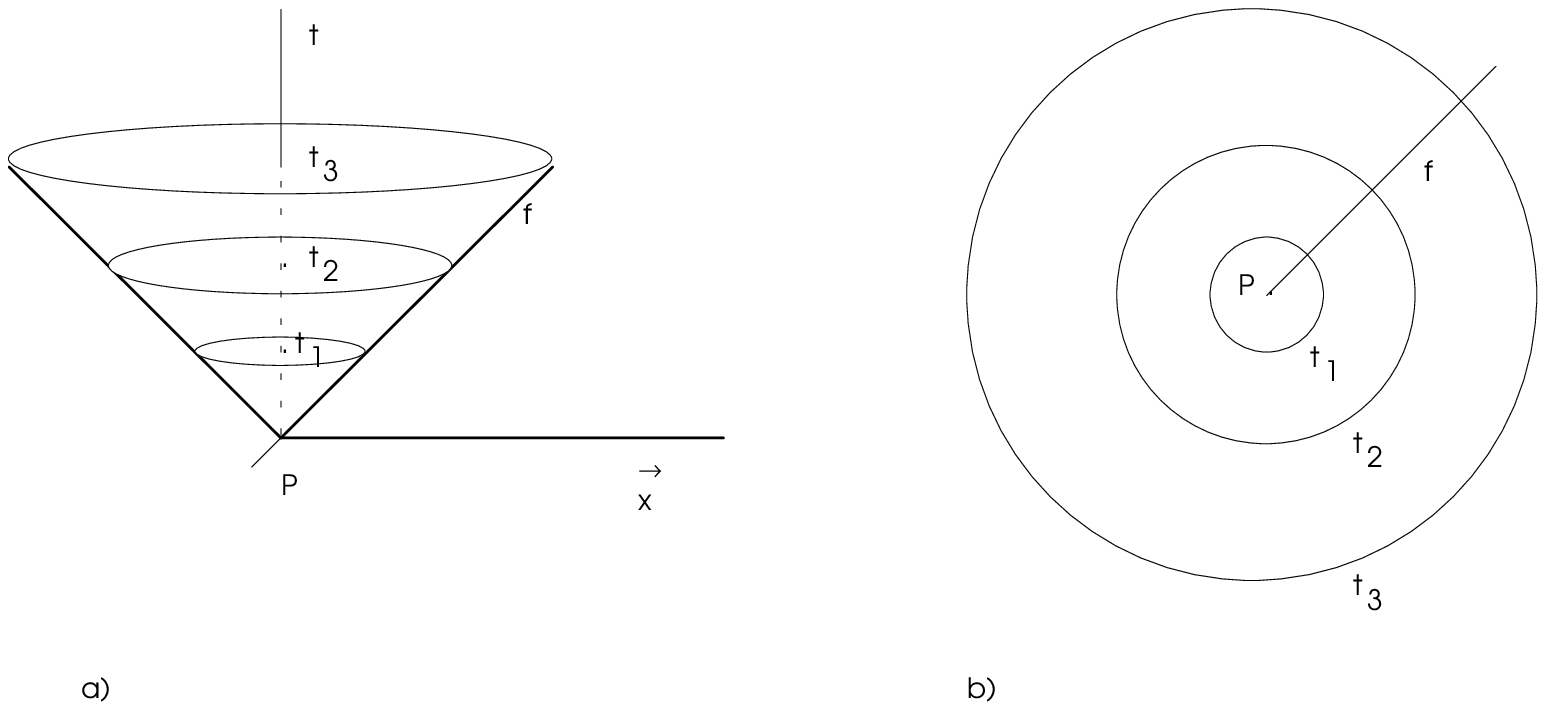}
\vglue-11cm}

\parbox[]{15.0cm}{Fig. 3. The front of a travelling spherical wave at three instants of time: (a) a spacetime  
diagram; (b) a three-space diagram. $f$ is a cone generator.}
\vglue1cm

This definition (\ref{Af}) would not make any sense if the point character (discrete and localized) of $A_{f}$ could not be sustained during its time evolution governed by its wave equation. Basing on the Huygens's principle one could erroneously think that it would not, but it is remarkable that it remains as a point-like field \cite {hep-th/9712069} as it propagates.  
Conversely, we have that 
\be
\label{s}
A(x,\tau)=\frac{1}{4\pi}\int d^{2}\Omega_{f}A_{f}(x,\tau),
\ee
where the integral represents the sum over all $f$ directions on the cone (\ref{1}). $4\pi$ is a normalization factor, $4\pi=\int d^{2}\Omega_{f}.$

Let us consider, just for fixing the idea, the electromagnetic theory as an example. Thus, $A(x,\tau)$ is the four-vector potential of an electromagnetic radiation (for simplicity) field. The physical interpretation associates $A_{f}(x,\tau)$, a point-perturbation propagating along the lightcone generator $f,$ with a physical photon - we call it a classical  photon - and $A(x,\tau)$, the standard continuous field, to the effect of the classical photon smeared on the lightcone spacetime. It is worthwhile to remind and to underline here the physical distinction \cite{hep-th/9712069} between $A_{f}(x,\tau)$ and $A(x,\tau)$. They do not represent  equivalent physical descriptions. $A_{f}(x,\tau)$  corresponds to a single real physical photon with $f$ being its four-vector velocity and with transverse electromagnetic fields, while $A(x,\tau),$  due to the smearing process (\ref{s}), corresponds to a continuous distribution of fictitious unphysical photons with longitudinal electromagnetic field. For retrieving the standard field, defined over the lightcone, of the standard formalism we necessarily have to introduce these fictitious longitudinal photons. This result matches with known theorems from field theory \cite{Strochi} and explains all the unreasonable difficulties \cite{Gupta} we have on quantizing the Maxwell field $A(x,\tau)$, as the photon is supposedly the simplest Nature's elementary object, assuming the inexistence of elementary scalar fields.  With $A_{f}(x,\tau)$, both the Lienard-Wiechert solutions, the advanced and the retarded, can be interpreted in terms of creation and annihilation of classical photons, without any problems of causality violation. See the Figure 4.
\mbox{} 
\begin{minipage}[]{5.0cm} 
\parbox[b]{5.0cm}{
\begin{figure}
\vglue2cm
\epsfxsize=100pt
\epsfbox{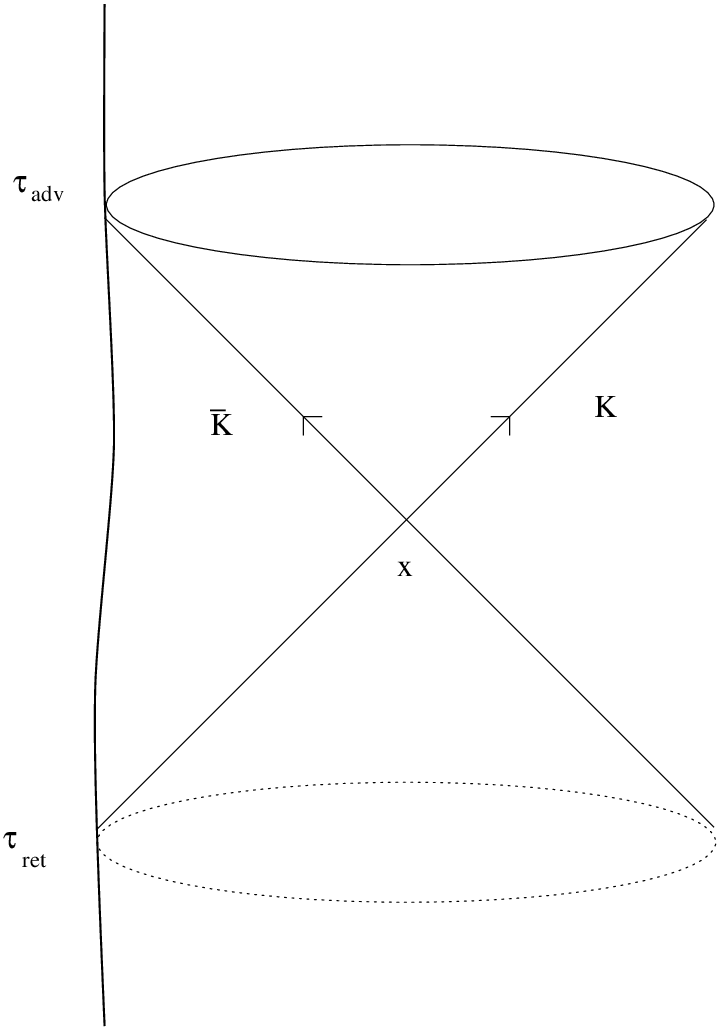}
\vglue-4cm
\end{figure}}\\
\end{minipage}\hfill
\hfill 
\mbox{}

\hspace{8.0cm}
\vglue-1cm
\begin{minipage}[]{7.0cm}\hglue7.0cm
\parbox[b]{7.0cm}
{\vglue-17cm \hspace{5.0cm}\parbox[t]{7.0cm}{
Fig.4. { 
{Creation an annihilation of particle in classical physics as a new interpretation of the LWS. At x there are two (classical) photons. One, created in the past by J, at $\tau_{ret},$ and propagating along the light cone generator K.  J is its source. The other one, propagating along ${\bar{K}}$, will be absorbed in the future by J, at $\tau_{adv}.$ J is its sink. Both are retarded and point-like solutions.}}}}
\vglue3cm
\end{minipage}

\vglue2cm

Another remarkable distinction, that will be also highly relevant for the gravitational field, is that $A_{f}(x,\tau)$ is a finite pointwise field while $A(x,\tau)$ has a singularity \cite {hep-th/9712069} introduced by the smearing process (\ref{s}). The reason for this great difference is that a cone is not a complete manifold as it is singular at its vertex. An extended field defined with support on a cone hypersurface is necessarily a singular field at the cone vertex, regardless its symmetries. The extended-field singularity just reflects the singularity of its support manifold. It is not a physical artifact. Figure 5 shows the relationship between the fields $A_{f}$ and $A$ for a process involving the emission of a single physical photon $A_{f}$; $A$ here is its space average.

\vglue1.5cm

\parbox[]{7.5cm}{
\begin{figure}
\vglue-7.5cm
\epsfxsize=400pt
\epsfbox{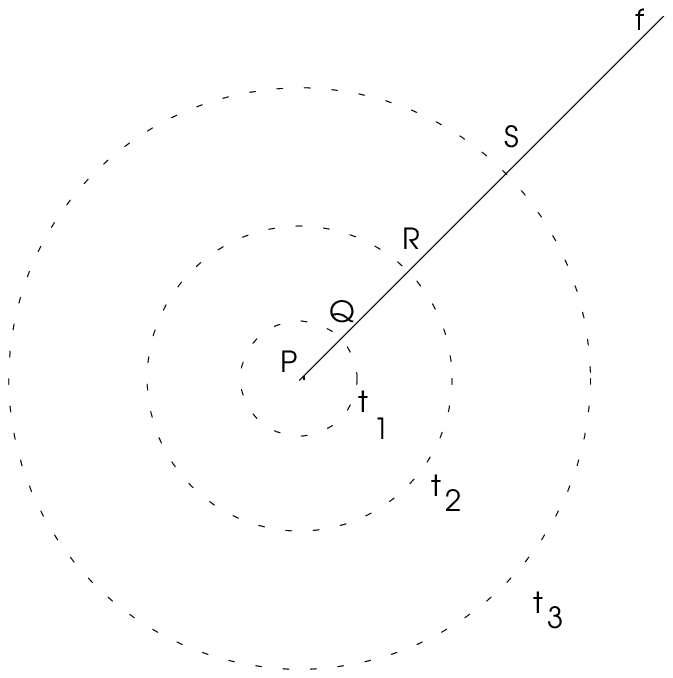}
\vglue-10cm
\end{figure}
\vglue0.5cm}
\mbox{}
\hfill
\hspace{6.0cm}
\parbox[]{7.5cm}{\vglue-2.5cm Fig. 5. A very low intensity light with just one photon. The three dotted circles represent the expanding Maxwell field for this light, at three instants of time. They transmit a false idea of isotropy.  The straight line PQRS\dots is the fibre $f,$ a lightcone generator tangent to $f^{\mu}.$ The points Q, R, and S, intersections of the fibre $f$ with the three dotted circles, are the single emitted classical photon $A_{f}$ at three instants of time.}\\ \mbox{}

\vglue1cm

\noindent The derivatives of $A_{f}(x,\tau),$ allowed by the constraint (\ref{lambda}), are the directional derivatives along $f,$ which with the use of (\ref{del}) we write as
\be
\label{fd}
\partial_{\mu}A_{f}=(\frac{\partial }{\partial x^{\mu}}+\frac{\partial \tau}{\partial x^{\mu}}\frac{\partial}{\partial \tau})A(x,\tau){\Big |} _{dx.\Lambda^{f}.dx=0}={\Big(}\frac{\partial }{\partial x^{\mu}}-f_{\mu}\frac{\partial}{\partial \tau}{\Big)}A_{f}\equiv\nabla_{\mu} A_{f}.
\ee
With $\nabla$ replacing $\partial$ for taking care of the constraint (\ref{lambda}), the propertime $\tau$ can be treated as a fifth independent  coordinate. 

\noindent The field equation for a massless field defined on a lightcone generator $f$ is, consequently,
\be
\label{wef}
\eta^{\mu\nu}\nabla_{\mu}\nabla_{\nu}A_{f}(x,\tau)=J(x,\tau),
\ee
or, explicitly
\be
\label{wef'}
(\eta^{\mu\nu}\partial_{\mu}\partial_{\nu}-2f^{\mu}\partial_{\mu}\partial_{\tau})A_{f}(x,\tau)=J(x,\tau),
\ee
as $f^{2}=0$. $J$ is its point-source four-vector current.\\ 
An integration over the $f$ degrees of freedom in (\ref{wef}) reproduces, with the use of (\ref{s}), the usual wave equation of the standard formalism, \be
\l{10'}
\eta^{\mu\nu}\partial_{\mu}\partial_{\nu} A(x,\tau)= J(x,\tau),
\ee
 as $\int d^{2}\Omega_{f}f^{\mu}\partial_{\mu}\partial_{\tau}A_{f}(x,\tau)=0$ because \cite{hep-th/9708066} $A_{f}(x,\tau)=A_{-f}(x,\tau)$. The standard formalism is retrieved from this $f$-formalism with the $A(x,\tau)$ as the average of $A_{f}(x,\tau)$, in the sense of (\ref{s}). 

\section{General Relativity with Extended Causality.}

We want to apply here this theory for the gravitational field in the Einstein's General Relativity with 
\be
g_{\mu\nu}^{f}=g_{\mu\nu}{\Big |} _{\Delta x.\Lambda^f.\Delta x=0}
\ee
\noindent The simplest way is just to write
\begin{equation}
\l{g}
g_{\alpha\beta}^{f}=\eta_{\alpha\beta}+H_{f}(x,\tau )f_\alpha f_\beta
\end{equation}
where $H_{f}(x,\tau )$ represents a local spacetime deformation produced by the presence of a single graviton propagating along a
D
straight line tangent to $f,\;f^{2}=0.$ The parameter $f$, we remind, is a constant four-vector, which expresses the graviton freedom as it freely propagates up to the point where it suffers an interaction (it is absorbed). The extended causality describes the straight-line motion (on a Minkowski spacetime) of a free point-field between two consecutive interactions of its sources; all sources and fields are point-like objects; all interactions are discrete and localized at a point and there is no place for self-interactions. This is just a consequence of $f$ being constant!  The Einstein's formalim remains diffeomorphism invariant. A flat background in this approach just represents the absence of any interaction, of any ``quantum" (discrete) of interaction. \\ From
$g^{\alpha\beta}_{f}g^{f}_{\beta\mu}=\delta^{\alpha}_{\mu}$ we have 
\be
\l{gup}
g^{\alpha\beta}_{f}=\eta^{\alpha\beta}-H_{f}(x,\tau )\,\eta^{\alpha\mu}f_{\mu}\,\eta^{\beta\nu}f_{\nu}.
\ee
As $f^{\mu}=: g_{f}^{\mu\nu}f_{\nu}=\eta^{\mu\nu}f_{\nu},$ because of $f^{2}=0$, we can write $g^{\alpha\beta}_{f}=\eta^{\alpha\beta}-H_{f}(x,\tau )f^{\alpha}f^{\beta}.$  Observe that $g^{\alpha\beta}_{f}$ and $g_{\alpha\beta}^{f}$ are both bi-linear on $f$. This enormous simplification --the absence of non-linearity-- is exclusively a consequence of (\ref{g}) and of $f$ being lightlike. They are justified with the classical vision of $g_{f}$ as a point-field describing a graviton, freely propagating with the velocity of light. It is important to remark that (\ref{g}) does not imply any kind of weak field approximation: $H_{f}(x,\tau )f_\alpha f_\beta$ is equal to $g_{\alpha\beta}^{f}-\eta_{\alpha\beta}$ whichever be $g_{\alpha\beta}^{f}$.
An immediate consequence of $f^{2}=0$ is that 
\be
det\; g_{\alpha\beta}^{f}=det\; \eta_{\alpha\beta}^{f}=-1,
\ee
which indicates that (\ref{g}) describes a singularity-free field.
Let us write $(\Gamma _{\beta \gamma }^\lambda )^f$ as the Christoffel symbols
restricted to the line $f,$ defined by 
\be
(\Gamma _{\beta \gamma }^\lambda
)^f=\frac 12g^{\lambda \sigma }_{f}(\nabla_{\gamma}g_{\beta
\sigma}^{f}+\nabla_{\beta}g_{\sigma \gamma}^{f}-\nabla_{\sigma}g_{\beta \gamma}^{f}).
\ee
For notation simplicity we will write $(\Gamma _{\beta \gamma }^\lambda )^f$ as just $\Gamma _{\beta \gamma }^\lambda$ without the index $f,$ and 
 $\nabla _\alpha H_{f}\;\corresponds\; H_\alpha $ and $\nabla _\alpha \nabla _\beta H_{f}\;\corresponds\; H_{\alpha \beta }.$ Then,
\begin{equation}
2\Gamma _{\alpha \beta }^\mu =f_\alpha f^\mu H_\beta +f^\mu f_\beta H_\alpha -f_\beta f_\alpha \eta^{\mu\nu}H_\nu ,
\end{equation}
and $2\Gamma ^\lambda\,\corresponds\, 2g^{\beta
\gamma }\Gamma _{\beta \gamma }^\lambda=f^\lambda f^{\alpha}H_{\alpha}\,\corresponds\, f^\lambda f.H\qquad$ So, the harmonic coordinate conditions ($\Gamma^{\lambda}=0$)  imply on
\begin{equation}
f.H =0.  \label{ddonder}
\end{equation}
We can have a better physical picture of the coordinate conditions (\ref{ddonder}) using (\ref{g},\ref{gup}) and $f^2=0$ to write
\be
2\Gamma^{\lambda}=\nabla_{\mu}g_{f}^{\lambda\mu}=0,
\ee
which shows the parallelism with the Lorentz gauge condition of the electromagnetic theory.
Its physical meaning is, however, best exposed in the case of solutions to the inhomogeneous field equations \cite{hep-th/9712069}: a constraint between the direction of emission (absorption) of a point-like field and the consequent changes in the state of motion of its source (sink). See eq. (\ref{fa}).

As a consequence of (\ref{ddonder}) we have $\Gamma _{\beta \lambda }^\lambda=0$ and
\begin{equation}
R_{\rho \alpha \beta \sigma }^f=\frac 12( f_\rho f_\beta H_{\sigma
\alpha }-f_\alpha f_\beta H_{\sigma \rho }-f_\rho f_\sigma H_{\beta \alpha
}+f_\alpha f_\sigma H_{\beta \rho }),  \label{TRiem}
\end{equation}
\begin{equation}
\l{Ricci}
R_{\rho \beta }^f=\frac 12f_\rho f_\beta \,\eta ^{\alpha \sigma }H_{\sigma
\alpha }\;\corresponds\;\frac 12f_\rho f_\beta \Box_fH,
\end{equation}
and 
\begin{equation}
R^f=g_f^{\rho \beta }R_{\rho \beta }^f\;\corresponds\; 0.  \label{EscR}
\end{equation}
For reasons of clarity and simplicity we will consider here, on this first work on this subject, just solutions to the homogeneous Einstein field equations, which, with (\ref{Ricci}) and (\ref{EscR}) are then reduced to
\begin{equation}
\l{box}
\Box _fH(x,\tau )=0.  \label{delH}
\end{equation}
A very simple equation indeed, a consequence of (\ref{g}) and of $f^{2}=0$. The light-like $f$ in (\ref{g}) eliminates all the intrinsic non-linearity  of General Relativity. But one should be warned again that only the inhomogeneous equations are completely meaningful in extended causality because the physical properties of the emitted (absorbed) field reflects the changes its emission (absorption) caused on the state of motion of its source (sink). The changes in the sources provide valuable informations about the field, like its angular momentum and its state of polarization.  We will not discuss any further these shortcomings as they are just consequences of a solution to an (homogeneous) equation without a source term. Nonetheless, this extremely simple, information depleted system, is reach enough to justify its presentation as a first introduction to the subject. It enlightens the physical significance of continuous solutions of the standard formalism, like the Schwarzschild metric for example.\\
\section{Discrete solution}
The most general solution to the equation (\ref{box}) can be obtained, for example, from a Fourier expansion 
\be
\l{ft}
H(x,\tau )=\int d^{5}p { H}(p)\;e^{i(p_{\mu}x^{\mu}+ p_{5}\tau)},
\ee 
with $x$ and $\tau$ treated as five independent variables. The simplest solution to (\ref{delH}) and (\ref{ft}) with $f^{2}=0,$ is $H(p)=\delta[(p_{\mu}-f_{\mu}p_{5})^{2}]$ or, for mathematical convenience, \be
H(p)=2\chi\frac{|p.f|}{p.f}\delta[(p_{\mu}-f_{\mu}p_{5})^{2}]=\frac{\chi}{p.f}\delta(p_{5}-\frac{p^{2}}{2p.f}),
\ee
where $\chi$ is a constant. Then we have that 
\be
\l{pr5}
H(x,\tau )=\chi\int d^{4}p\frac{e^{i(p_{\mu}x^{\mu}+ \frac{p^{2}}{2p.f}\tau)}}{p.f}.
\ee
It is crucial in this expression that one has $p.f$ in the integrand denominator instead of the $p^{2}$  that one would have in the usual local-causality formalism, which would give origin to a $\frac1r$-dependence and, therefore, a metric with a singularity on $r=0.$ The extended causality, with its anisotropy determined by the existence of a graviton, allows the replacement of $p^{2}$ by $p.f$ This radically changes the nature and characteristics of the theory.

As we have observed the integrand of (\ref{pr5}) has a singularity at $p.f=0$ but also the exponent in the integrand is defined at this point only if
\be
\label{p2f}
p^{2}{\Big |}_{p.f=0}=0.
\ee
This implies on a system of two simultaneous equations
\be
\label{p2}
p^{2}=p_{\hbox{\tiny T}}^{2}+p_{\hbox{\tiny L}}^{2}-p_{4}^{2}=0,
\ee
and 
\be
\label{pf}
p.f=p_{\hbox{\tiny L}}|{\vec f}|-p_{4}f_{4}=0,
\ee
where the subindexes {\tiny L} and {\tiny T} stand, respectively, for
longitudinal and
transversal with respect to the space part $\stackrel{\rightarrow}{f}$
of $f.$\\
As $f^{2}=0$ we may write $\epsilon=\frac{|{\vec f}|}{f_{4}}=\pm1$, so that (\ref{pf}) becomes $p_{4}=\epsilon p_{\hbox{\tiny L}}.$ Equation (\ref{p2}) is, consequently, equivalent to
\be
\label{pt}
p_{\hbox{\tiny T}}=0.
\ee
The conditions (\ref{p2}) and (\ref{pt}) are full of physical significance: the first one requires a massless field and the second one implies that $\Delta x_{\hbox{\tiny T}}=0$. Only the $x_{\hbox{\tiny L}},$ that is, the longitudinal coordinate, participates in the system evolution \cite{hep-th/9708066}.  So, one can see in anticipation, that the field $H_{f}$ only propagates along the  fibre $f.$ \\
The equation (\ref{pr5}), as a consequence of (\ref{pt}),  is reduced to 
\be
\label{pr6}
H_{f}(x,\tau)=\chi\int \frac{dp_{\hbox{\tiny L}}dp_{4}}{2f_{4}(p_{4}-\epsilon p_{\hbox{\tiny L}})}e^{i(p_{\hbox{\tiny L}}x_{\hbox{\tiny L}}+p_{4}t+\frac{p_{4}+\epsilon p_{\hbox{\tiny L}}}{2f_{4}}\tau)},
\ee
after having $\frac{p^{2}}{2p.f}$ re-written as 
\be
\frac{ p_{\hbox{\tiny L}}^{2}-(p_{4})^{2}}{2p.f}=\frac{(\epsilon p_{\hbox{\tiny L}}+p_{4})(\epsilon p_{\hbox{\tiny L}}-p_{4})}{2(\epsilon p_{\hbox{\tiny L}}-p_{4})f_{4}}=\frac{\epsilon p_{\hbox{\tiny L}}+p_{4}}{2f_{4}}.
\ee
Now we make explicit the integration on the coordinate $p_{4}$,
\be
\label{pr7}
H_{f}(x,\tau)=\lim_{\varepsilon\to0}\chi\int dp_{\hbox{\tiny L}}dp_{4}\frac{e^{i(p_{\hbox{\tiny L}}x_{\hbox{\tiny L}}+p_{4}t+\frac{p_{4}+\epsilon p_{\hbox{\tiny L}}}{2f_{4}}\tau)}}{2f_{4}(p_{4}-\epsilon p_{\hbox{\tiny L}}\pm i\varepsilon)},
\ee
which produces
\be
\label{pr8}
H(x,\tau)_{f}=2\pi ia\chi\theta[a(t+\frac{\tau}{2f_{4}})]\int \frac{dp_{\hbox{\tiny L}}}{2f_{4}}e^{ip_{\hbox{\tiny L}}(x_{\hbox{\tiny L}}+\epsilon t+\frac{\epsilon}{f_{4}} \tau)},
\ee
where $a$ stands for $\pm1$, a sign that comes from the choice of the contour in a Cauchy integral, i.e. the sign of $\pm i\varepsilon$. The signs of $a=\pm1$ are connected \cite{hep-th/9708066}, respectively, to the creation and annihilation of $g_{f}(x,\tau);$ and $\theta(t)$ is the step function ($\theta(t\ge0)=1;\;\theta(t<0)=0$). On the other hand
\be
\label{d}
\frac{1}{2\pi}\int \frac{dp_{\hbox{\tiny L}}}{f_{4}}e^{ip_{\hbox{\tiny L}}(x_{\hbox{\tiny L}}+\epsilon t+\frac{\epsilon} {f_{4}}\tau)}=\frac{1}{2\pi\epsilon}\int \frac{\epsilon dp_{\hbox{\tiny L}}}{f_{4}}e^{i\frac{p_{\hbox{\tiny L}}}{f_{4}}\epsilon(f_{\hbox{\tiny L}}x_{\hbox{\tiny L}}+f_{4} t+\tau)}=\frac{1}{\epsilon}\delta(\tau+f.x),
\ee
and therefore, we have for (\ref{pr8})
\be
\label{pr8'}
H(x,\tau)_{f}=\frac{ia(2\pi)^{2}}{\epsilon}\chi\theta[a(t+\frac{\tau}{2f_{4}})]\delta(\tau+f.x),
\ee
which, after some simple algebra and a redefinition of the constant $\chi$, may be written as
\be
H(x,\tau)_f=\chi\theta (at)\delta (\tau+f\cdot x). 
\ee
Thus, with $\tau=0$ accounting for the massless field, and $t>0,$ because we are considering only the emitted field ($a=+1$) and not the absorbed one, the metric (\ref{g}) becomes
\begin{equation}
g_{\alpha \beta }^{f}=\eta _{\alpha \beta }+\chi \delta
(f\cdot \Delta x)f_\alpha f_\beta.  \label{solH}
\end{equation}
In spherical coordinates we have 
\be
\l{gtf}
g^{f}_{\alpha\beta}(t,r,\theta,\varphi)=\cases{\eta_{\alpha\beta},& for $\theta\ne\theta_{f},\; \varphi\ne\varphi_{f}$;\cr
\eta_{\alpha\beta}+Hf_{\alpha}f_{\beta}& for $\theta=\theta_{f},\; \varphi=\varphi_{f}$,\cr}
\ee
or explicitly 
\begin{equation}
\label{recSch}
\text{ }g^{f}_{\alpha \beta }(t,r,\theta ,\varphi )=\pmatrix{
-1 & 0 & 0 & 0 \cr 
 0 & 1 & 0 & 0 \cr 
 0 & 0 & r^2 & 0 \cr 
 0 & 0 & 0 & r^2\sin^2\theta\cr} +\chi\delta(f.\Delta x)\pmatrix{
{ f}_0^2 & { f}_0{ f}_r & 0 & 0 \cr 
{ f}_r{ f}_0 & { f}_r^2 & 0 & 0 \cr 
0 & 0 & 0 & 0 \cr 
0 & 0 & 0 & 0\cr},  
\end{equation}
for $\theta=\theta_{f},\; \varphi=\varphi_{f}$
with  $f_{\mu}=(f_0,\ f_r,0,0),$ and with $\theta_{f}$ and $\varphi _{f} ,$ defining the space direction  $\vec{f}$ of $f$. The metric $\,\,\ g^{f}_{\alpha \beta }$
represents a single, let's say, ``classical quantum" of gravity  propagating along a line $f$  and
observed as an event $(t,r,\theta _{f},\varphi _{f})$ at the probe mass. Let us, in an abuse of language,  call it the graviton on the fibre $f,$ for shortness. 
\section{Retrieving the Schwarzschild field}
The  presence of a ``graviton" on the fibre $f$  breaks the otherwise spherical symmetry in (\ref{gtf},\ref{recSch}).  It is not, of course, a static solution. There is no static solution in an extended causality formalism. As we will see, the observed (gravitational, like the electromagnetic) static fields are just average fields, apparently static as a consequence of the large number of quanta exchanged and of the inertial limitations of our measuring apparatus. From this discrete, localized and singularity-free solution $g_{\alpha\beta}^{f}$ on the lightcone-generator $f$ we can recover the standard continuous and distributed solutions $g_{\alpha\beta}$ with just an integration over the $f$-parameter. In order to obtain the standard  continuous  solution  the single physical graviton $g^{{ f}}_{\mu\nu}$ must be replaced by a continuous distribution of fictitious (non-physical) gravitons $g^{f'}_{\mu\nu}$, each one still propagating on the same fibre $f$. What distinguishes the physical graviton from the fictitious ones is that $f'$ does not satisfy the gauge condition (\ref{ddonder});  only ${ f}$ corresponds to the field four-velocity, and so, only ${\vec f}$ is collinear to ${\vec x}$, the direction of propagation of the gravitons. This is the exact analogous to what happens in the discretization of the electromagnetic field \cite{hep-th/9712069}. The continuous solution so obtained is determined by its chosen symmetry. Let us choose an spherically symmetric (on $f'$) distribution of $g^{f'}_{\mu\nu}$ so that we have

\begin{equation}
g_{\alpha \beta }(x,\tau)=\frac 1{4\pi }\int d^2\Omega _{f'}\ g_{\alpha \beta }^{
f'}(x,\tau).  \label{integab}
\end{equation}
We choose
\be
\l{f4}
f'^{4}=f'_{r}=1,
\ee
breaking then the, up to here, explicit Lorentz covariance. The physical meaning of this choice would also be better appreciated in a context of non-homogeneous field equations. In order to understand it, it is worthwhile to make a brief regression\cite{hep-th/9712069} on the extended causality condition, $\Delta\tau+f.\Delta x=0$, for $\Delta\tau=0$, that is, for $f.(x-z(\tau))=0$, where $z(\tau)$ is the field source worldline, parameterized by $\tau.$ Thus, $\nabla_{\beta}f.\Delta x=0$ implies on $f_{\beta}(1+f.V)=0$ or
\be
\l{fV}
f'.V{\Big |}_{\tau_{ret}}=-1,
\ee
with $V\,\corresponds\,\frac{dz}{d\tau},$ which represents a constraint between the direction of the emitted graviton and the instantaneous velocity V of its source at the emission time. So, 
\be
f'^{4}{\Big|}_{{V=o}\atop{f.\Delta x=0}}=1,
\ee 
and the choice (\ref{f4}) means that we are in the source instantaneous rest frame. This condition (\ref{fV}) is essential to get the Schwarzschild solution in its standard form, that is in its singularity rest-frame.
So, we can write
\begin{equation}
g_{\alpha \beta }=\eta _{\alpha \beta }+\frac {\chi}{2\pi}\pmatrix{ 
1 & -1 & 0 & 0 \cr 
-1 & 1 & 0 & 0 \cr 
0 & 0 & 0 & 0 \cr 
0 & 0 & 0 & 0\cr}
\int d^2\Omega _{f'}\ \ \delta (f'\cdot x).
\end{equation}
Writing $f'.x=f'^{4}t-{\vec f}'.{\vec r}=t- r\cos\theta_{f'}$, the integration  $\int d^2\Omega _{f'}\ \delta (f'\cdot x)$  may be
written as
\begin{equation}
\l{r}
\int d\varphi _{f'}\ \sin\theta _{f'}\ d\theta_{f'}\ \delta (t-\mid\overrightarrow{r}\mid \cos \theta _{f'})=\frac{2\pi }r\int_{-1}^{1}d\cos
\theta _{f'}\,\delta (\cos \theta _{f'}-\frac{t}r)=
\cases{\frac{2\pi }r,&for $t\in[0,r]$;\cr
0,&for $t\notin[0,r]$.\cr}
\end{equation}
Here it becomes evident why it is necessary to introduce a distribution of fictitious $g^{f'}_{\mu\nu}$ replacing the physical $g^{f}_{\mu\nu}$ because $\theta_{f}$ is, by definition, null and the integration on $\theta_{f'}$ is essential for getting the factor $\frac{1}{r}$ in (\ref{r}).
The condition on $t$ in (\ref{r}) means that the deformation on the flat spacetime that we are associating to a graviton is not null as far as $t$ is smaller, or at least, equal to $\frac rc$, the time that the graviton, after being emitted by the source at the origin, takes to reach the probe mass at $(t,r,\theta_{f},\phi_{f})$, where it is absorbed. This process is continued for $t>\frac rc$  by other gravitons subsequently emitted \cite{hep-th/9610145}. So, the large number of gravitons emitted (and absorbed) in any realistic experiment transmit the idea of continuity and of a static field. Thus we can write $$ds^2=-(1-\frac \chi
r)\,dt^2+(1+\frac \chi r)\,dr^2-\frac{2\chi}{r}dtdr+r^2(d\theta ^2+\sin^2\theta
\,d\varphi^2)$$
\\A well-known \cite{Weinberg} simple coordinate transformation leads it to
\begin{equation}
ds^2=-(1-\frac \chi
r)\,dt^2+(1-\frac \chi r)^{-1}\,dr^2+r^2(d\theta^2+\sin^2\theta
\,d\varphi^2).  \label{campoSch}
\end{equation}
We recognize (\ref{campoSch}) as the Schwarzschild metric. Its global validity is restricted by (\ref{gtf}) being a vacuum solution. This approach requires that the field source be treated as a set of pointlike sources whichthe imposed spherical symmetry in (50) reduces to the equivalent to a single point-like source. It is a consequence of the assumed isotropy in the distribution of fictitious gravitons; other distinct symmetries, of course, generate other distinct metrics.
The probe mass, wherever be it placed, detects the Schwarzschild field on the space around the coordinate origin $r=0$. Eq. (\ref{campoSch}) describes an average gravitational interaction between the test-body and the point-source at $r=0.$ We leave possible alternative interpretations to be discussed elsewhere. 

We started with a theory for the radiation field to find out, {\it a posteriori}, that it applies to static fields too. This could be a pleasant surprise if it hadn't already\cite{hep-th/9712069} happened to the electromagnetic field. Actually it presents a new vision of an static field as a radiation in an appropriate limit where the field discreteness is smothered out. So, we could say that, in this context,  what has not  been detected yet by our gravitational wave detectors, is just a coherent or a ``low"-frequency gravitational radiation.

\section{Conclusions}

We have shown, in previous works \cite{hep-th/9610028,hep-th/9708066}, that the problems of classical field theories with singularities,  divergencies, and difficulties of quantization are consequences of being defined with support on the lightcone; then their fields are not the real fundamental ones but just their effective averages. The actually fundamental fields must be defined with support on the lightcone generators. This corresponds to adopting extended, instead of local, causality. The Maxwell theory of electromagnetism has been shown \cite{hep-th/9712069} to be free of these problems when formulated on the lightcone generator, that is, in terms of finite and discrete point-like fields (classical photons). The extended causality gives a better description of electromagnetism; the standard formalism with all of its known problems is recuperated when the photon fields are replaced by continuous fields defined by the photon effective averages on the lightcone. An important message then is that electromagnetic field singularities are not real physical objects but just artificial consequences of using an inappropriate formalism. The remarkable in the present work is that all these considerations on the electromagnetic field are now repeated for the gravitational field of the General Theory of Relativity; their similarities are greatly enhanced.
In General Relativity, like in Electrodynamics \cite{hep-th/9712069}, the standard continuous field can be retrieved from the discrete one through an averaging process that requires the inclusion of fictitious unphysical fields necessarily. In Electrodynamics these unphysical fields are the responsible for the complications on an otherwise simple quantization process; one may assume that in General Relativity they make this quantization impossible. In both theories singularities are just consequences of the averaging process, of using these averages as if they were the actual fundamental fields.

 The Schwarzschild metric, a static spherically symmetric field, can be seen as the average effect of the flux of discrete point-like fields. A classical graviton is a constant point-like disturbance on the spacetime fabric, propagating without a change on itself. It propagates on a background Minkowski spacetime reflecting the basic assumption that all interactions have a fundamental quantum (discrete) nature. A flat background in this approach signals the absence of any quantum of interaction. Its time independence and its singularity, a really not physical object, are consequences of taking an average by the fundamental field.  This is quite a change and, certainly, of no easy acceptation as it goes against the prevalent trend of seeing the field singularities as real physical objects, and the continuous metric field as a true physical representation of the world geometry (not just an approximation), notwithstanding the unsurmountable difficulties that this implies on having a quantum theory for gravity.
\acknowledgments
R. N. Silveira acknowledges a grant from CAPES for writing his M.Sc. dissertation.

\newpage
\begin{center}
List of figure captions.
\end{center}
\begin{enumerate}
\item Fig.1. The relation $\Delta\tau^2=-\Delta x^{2},$ a causality constraint, is seen as a restriction of access to regions of spacetime. It defines a three-dimension cone which is the spacetime available to a point physical object at the cone vertex. 
\item Fig.2. The usual interpretation of the Lienard-Wiechert solutions. By the point x passe two spherical waves: the retarded one, created in the past $\tau_{ret}$, and the advanced one, created in the future $\tau_{adv}.$ J is the source of both.
\item Fig. 3. The front of a travelling spherical wave at three instants of time: (a) a spacetime diagram; (b) a three-space diagram. $f$ is a cone generator.
\item Creation an annihilation of particle in classical physics as a new interpretation of the LWS. At x there are two (classical) photons. One, created in the past by J, at $\tau_{ret},$ and propagating along the light cone generator K.  J is its source. The other one, propagating along ${\bar{K}}$, will be absorbed in the future by J, at $\tau_{adv}.$ J is its sink. Both are retarded and point-like solutions.
\item Fig. 4. Creation an annihilation of particle in classical physics as a new interpretation of the LWS. At x there are two (classical) photons. One, created in the past by J, at $\tau_{ret},$ and propagating along the light cone generator K.  J is its source. The other one, propagating along ${\bar{K}}$, will be absorbed in the future by J, at $\tau_{adv}.$ J is its sink. Both are retarded and point-like solutions.
\item Fig. 5. A very low intensity light with just one photon. The three dotted circles represent the expanding Maxwell field for this light, at three instants of time. They transmit a false idea of isotropy.  The straight line PQRS\dots is the fibre $f,$ a lightcone generator tangent to $f^{\mu}.$ The points Q, R, and S, intersections of the fibre $f$ with the three dotted circles, are the single emitted classical photon $A_{f}$ at three instants of time.
\end{enumerate}
\end{document}